\documentclass[prl,preprint,showpacs,preprintnumbers]{revtex4}
\usepackage{amsmath}
\usepackage{graphicx}
\usepackage{epsfig}
\usepackage{dcolumn}
\usepackage{bm}

\input{tcilatex}

\begin{document}

\title{Physically-motivated dynamical algorithms for the graph isomorphism
problem}
\author{Shiue-yuan Shiau}
\author{Robert Joynt}
\author{S.N. Coppersmith}
\affiliation{Department of Physics, University of Wisconsin, Madison, Wisconsin 53706}

\begin{abstract}
We investigate classical and quantum physics-based algorithms for
solving the graph isomorphism problem. Our work integrates and
extends previous work by Gudkov et al. (cond-mat/0209112) and by
Rudolph (quant-ph/0206068). Gudkov et al. propose an algorithm
intended to solve the graph isomorphism problem in polynomial time
by mimicking a classical dynamical many-particle process. We show
that this algorithm fails to distinguish pairs of non-isomorphic
strongly regular graphs, thus providing an infinite class of
counterexamples. We also show that the simplest quantum
generalization of the algorithm also fails. However, by combining
Gudkov et al.'s algorithm with a construction proposed by Rudoph
in which one examines a graph describing the dynamics of two
particles on the original graph, we find an algorithm that
successfully distinguishes all pairs of non-isomorphic strongly
regular graphs that we tested (with up to 29 vertices).
\end{abstract}

\maketitle

\textbf{Introduction. \ }The graph isomorphism problem plays a central role
in the theory of computational complexity and has importance in physics and
chemistry as well\cite{kobler93,fortin96}. \ A graph is a set of $N$ points,
or vertices, and a set of edges, or unordered pairs of those points. \ If
two graphs differ only in the labelling of their points, then we say they
are isomorphic, otherwise they are non-isomorphic. \ The graph isomorphism
problem is to determine whether or not there is an algorithm that runs in
polynomial time $(t$ $\sim N^{x},$ with $x$ independent of $N$) that
distinguishes non-isomorphic pairs with certainty. \ The efficient
enumeration of possible distinct atomic clusters of size $N$, in which a
cluster is defined by the bonds between its atoms, is the same problem in
another guise\cite{liu91}.

Typical instances of graph isomorphism (GI) can be solved in polynomial time
because two randomly chosen graphs with identical numbers of vertices and
edges typically have different degree and eigenvalue distributions.
Moreover, GI can be solved efficiently for restricted classes of graphs,
such as trees\cite{hopcroft71}, planar graphs\cite{hopcroft74}, graphs with
bounded degree\cite{luks82,hoffman82}, bounded eigenvalue multiplicity\cite%
{babai82}, and bounded average genus\cite{chen94}. However, no algorithm for
solving GI for all graphs is presently known; the best existing upper bound
is $\exp \sqrt{cN\log N}$\cite{babai83}. There is evidence that indicates
that GI is not NP-complete: first, counting the number of isomorphisms is
reducible to the decisional version of the problem\cite{mathon79}, unlike
the situation for all known NP-complete problems, and second, if GI were
NP-complete then certain complexity classes that are believed to be
different would be identical\cite{boppana87,schoning88}. Thus, it is
believed that either GI is in P or else it is in the class of problems that
are neither in P nor are NP-complete\cite{factorization_footnote}

One way to solve GI is to solve the hidden subgroup problem for the
permutation group. Unfortunately, though the hidden subgroup problem for
abelian groups can be solved efficiently\cite{shor97}, no efficient
algorithm for solving the hidden subgroup problem for the permutation group
is known~\cite{grigni01,hallgren03}.

In this paper we investigate classical and quantum approaches to solving the
graph isomorphism problem that are motivated by physical processes.\cite%
{adiabatic_footnote} Our work combines and extends ideas in Gudkov \textit{%
et al.}\cite{gudkov02} and by Rudolph\cite{rudolph02} for attacking the GI
problem using algorithms based on physical intuition. We show that the
interesting classical dynamical algorithm proposed by Gudkov \textit{et al.}
fails to distinguish an infinite set of pairs of non-isomorphic graphs, and
thus does not solve the GI problem in polynomial time. We trace this failure
to certain algebraic properties of these particular pairs of graphs and show
that the simplest quantum generalizations of the Gudkov \textit{et al}.
algorithm also fails to distinguish these pairs of graphs. However, an
algorithm obtained by combining the basic idea of the Gudkov \textit{et al}.
algorithm with a construction proposed by Rudolph\cite{rudolph02} does
distinguish all pairs of graphs that we have examined, including many with
the same eigenvalue spectra. 

The detailed statement of the GI problem is as follows. \ We are given two
graphs. \ The first is a set of $N$ vertices $\left\{
v_{1},v_{2},...v_{N}\right\} ,$ together with a set of edges, or unordered
pairs $\left\{ e_{1},e_{2},...\right\} $ connecting pairs of these vertices,
while the second graph is a set of vertices $\left\{ v_{1}^{\prime
},v_{2}^{\prime },...v_{N}^{\prime }\right\} ,$ and a set of edges $\left\{
e_{1}^{\prime },e_{2}^{\prime },...\right\} $. \ Each $e_{i}$ is associated
with a pair $\left\{ v_{a},v_{b}\right\} $, and each $e_{i}^{\prime }$ is
associated with a pair $\left\{ v_{a}^{\prime },v_{b}^{\prime }\right\} .$ \
We wish to determine whether there exists a permutation $\mathcal{P}$ of the
$v_{i}$ such that the set of pairs $\left\{ \mathcal{P}v_{a},\mathcal{P}%
v_{b}\right\} $ is identical with the set of pairs $\left\{ v_{a}^{\prime
},v_{b}^{\prime }\right\} .$

An equivalent but more useful formulation for our purposes is to represent
each graph by its adjacency matrix $A$. \ $A_{ab}$ is an $N\times N$ matrix
such that $A_{ab}=1$ if there is a pair $e=\{v_{a},v_{b}\}$ and $A_{ab}=0$
otherwise. \ Two graphs are isomorphic if and only if there exists a
permutation matrix $P$ such that $A^{\prime }=PAP^{-1}.$

\ \ \ \ \ \textbf{Classical algorithm. \ }In the Gudkov \textit{et al}.
approach to this problem\cite{gudkov02}, each vertex of the graph is mapped
into a point in an $N-$dimensional vector space. \ The vertex $v_{1}$ sits
initially at the point $\vec{r}_{1}=(1,0,0,...,0),$ $v_{2}$ at $\vec{r}%
_{2}=(0,1,0,...0),$ and so on. \ These are the vertices of an $(N-1)-$%
dimensional simplex. \ We now view these as mass-points with forces acting
pairwise among them.\ There is one force law if two particles are connected
by an edge and a different force law if they are not. \ The forces are
derived from potentials $U_{1}$ and $U_{2}$ that depend only on the
distances between the points in the pair. \ From the initial configuration,
the particles move in time according to the relaxational equations of motion
\
\begin{equation*}
\mu \frac{d\vec{r}_{a}(t)}{dt}=\vec{F}_{a},
\end{equation*}%
where
\begin{equation*}
\vec{F}_{a}=-\nabla _{\vec{r}_{a}}U(\vec{r}_{1},\vec{r}_{2},...)
\end{equation*}%
and%
\begin{equation}
U=\sum_{a>b}A_{ab}U_{1}(\left\vert \vec{r}_{a}-\vec{r}_{b}\right\vert
)+\sum_{a>b}(1-A_{ab})U_{2}(\left\vert \vec{r}_{a}-\vec{r}_{b}\right\vert ).
\label{eq:motion}
\end{equation}%
The motion may be computed by any convenient algorithm. \ After a time $T\,\
$the positions are given by $\vec{r}_{a}(T)$ for $a=1,2,...N.$ \ We now
compute the set of $N(N-1)/2$ distances $d_{ab}=\left\vert \vec{r}_{a}(T)-%
\vec{r}_{b}(T)\right\vert .$ \ Given a second graph, we compute $%
d_{ab}^{\prime }=\left\vert \vec{r}_{a}^{\prime }(T)-\vec{r}_{b}^{\prime
}(T)\right\vert $ using the same prescription. \ The sets $\left\{
d_{ab}\right\} $ and $\{d_{ab}^{\prime }\},$ being non-negative real
numbers, may be arranged in increasing order. \ If the resulting sets are
identical, then it is conjectured that the graphs are isomorphic. \ Note
that as long as $T$ is not an exponentially long time, this is a
polynomial-time algorithm. \ The sets $d_{ab}$ and $d_{ab}^{\prime }$ can be
computed, ordered, and compared efficiently. \ We wish to examine the claim
that if $\left\{ d_{ab}\right\} $ and $\{d_{ab}^{\prime }\}$ are the same up
to reordering, then their parent graphs are isomorphic. \ \

For purposes of clarity, we shall initially consider a slightly simpler
model, one in which a harmonic force acts only between particles connected
by edges:

\begin{equation*}
\widetilde{U}=-\mu \sum_{a>b}A_{ab}|\vec{r}_{a}-\vec{r}_{b}|^{2}/2.
\end{equation*}%
We then define an $N\times N$ matrix $X,$ where the $i$-th coordinate of the
$a$-th "particle"\ is denoted by $X_{ai}.$ Thus the above initial condition
can be rewritten as
\begin{equation*}
X_{ai}(t=0)=\delta _{ai},
\end{equation*}%
where $\delta _{ai}$ is the Kronecker symbol: $\delta _{ai}=1$ if $a=i$ and $%
\delta _{ai}=0$ otherwise. \ The equations of motion are
\begin{eqnarray*}
\frac{dX_{ai}}{dt} &=&\sum_{b}F_{ab}^{(i)}=\sum_{b}A_{ab}(X_{ai}-X_{bi})=-%
\sum_{b}A_{ab}X_{bi}+\sum_{b}\sum_{c}A_{ac}\delta _{ab}X_{bi} \\
&=&\sum_{b}L_{ab}X_{bi},
\end{eqnarray*}%
where%
\begin{equation*}
L_{ab}=\delta _{ab}\sum_{c}A_{ac}-A_{ab}=D_{ab}-A_{ab},
\end{equation*}%
is the Laplacian matrix. \ The diagonal matrix $D$ is the degree sequence
matrix: $D_{aa}$ is the number of edges incident to the vertex $a.$ \ Note
that $L$ is \ symmetric: $L=L^{T}.$ \ The algebraic isomorphism criterion
mentioned in the introduction may also be put in terms of $L:$ two graphs
defined by the Laplacian matrices $L$ and $L^{\prime }$ are isomorphic if
and only if there exists a permutation matrix $P$ such that%
\begin{equation*}
L^{\prime }=PLP^{T}.
\end{equation*}

The solution to the equation of motion for the particles defined by $%
\widetilde{U}$ is%
\begin{equation*}
X_{ai}(t)=\left( e^{Lt}\right) _{ab}X_{bi}(t=0)=\left( e^{Lt}\right) _{ai}
\end{equation*}%
so the final positions are
\begin{equation*}
X(T)=e^{LT},
\end{equation*}%
in a matrix notation. \ In order to compute the distances, we note that the
set of dot products between the position vectors may be written as:%
\begin{equation*}
S_{ab}=\sum_{i}X_{ai}X_{bi}
\end{equation*}%
which is the matrix%
\begin{equation*}
S=XX^{Transpose}=X^{2}=e^{2Lt}=1+2tL+\frac{\left( 2t\right) ^{2}}{2!}%
L^{2}+...
\end{equation*}%
Another graph would be characterized by a different dot product matrix%
\begin{equation*}
S^{\prime }=e^{2L^{\prime }t}.
\end{equation*}%
Since the $N^{2}$ distances $d_{ab}$ satisfy $%
d_{ab}^{2}=S_{aa}+S_{bb}-2S_{ab},$ comparing the set of numbers in the
matrices $S$ and $S^{\prime }$ is essentially the same as comparing the
distances. \ (We shall consider the relationship more carefully below.)

If the graphs are isomorphic, then clearly $S^{\prime }$ is a rearrangement
of $S:$%
\begin{equation*}
S^{\prime }=e^{2L^{\prime }T}=e^{2PLP^{-1}T}=Pe^{2Lt}P^{-1}=PSP^{-1}.
\end{equation*}%
The interesting question is whether the converse is also true.

\begin{figure}[t!]
\vspace{-3cm} \hspace{-8cm} \setlength{\unitlength}{0.8cm}
\begin{picture}(3,9)(-1.1,-0.5)
\put(1.5,1.5){\circle*{0.1} $5$} \put(0,0){\circle*{0.1}$4$}
\put(0,3){\circle*{0.1}$1$} \put(3,3){\circle*{0.1}$2$}
\put(3,0){\circle*{0.1}$3$}

\put(1.5,-0.5){$G$} \put(7.5,-0.5){$G^{\prime}$}

 \put(0,3){\line(1,-1){3}}
 \put(0,0){\line(1,1){3}}
\put(6,0){\circle*{0.1}$4$} \put(6,3){\circle*{0.1}$1$}
\put(7.5,1.5){\circle*{0.1}$5$} \put(9,0){\circle*{0.1}$3$}
\put(9,3){\circle*{0.1}$2$}
 \put(6,3){\line(1,0){3}}
 \put(6,0){\line(1,0){3}}
  \put(6,0){\line(0,1){3}}
 \put(9,0){\line(0,1){3}}

\end{picture}
\caption{Two isospectral graphs}
\label{fig:isographs}
\end{figure}
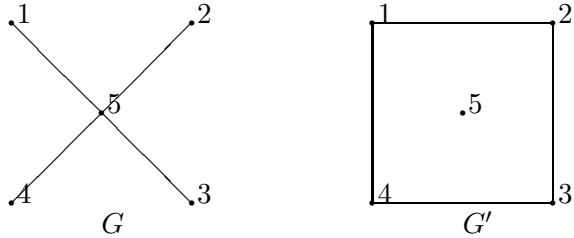

In fact there do exist many interesting graph pairs for which the algorithm
works. Because $S$ is most easily computed by diagonalizing the real
symmetric matrix $L,$ \ it is natural to ask whether pairs of non-isomorphic
but isospectral graphs can be distinguished by the method. \ Isospectral
graphs are those for which the eigenvalues of $A$ and $A^{\prime }$ are the
same. \ We have investigated this question for some small graphs that are
isospectral but not isomorphic. \ A simple illustrative pair is shown in
Fig.~\ref{fig:isographs}

We take the total time interval as $T=1$ and compute numerically the
dynamics of the simple harmonic model for $10$ steps of length $0.1$ using
the first-order Euler algorithm and finally obtain a normalized $X(T)$. The
sorted $d_{ab}^{2}$ and $d_{ab}^{2\prime }$ for two graphs $G$ and $%
G^{\prime }$ in Fig.~\ref{fig:isographs} are respectively
\begin{eqnarray}
d_{ab}^{2} &=&\left\{ [0]^{5},[0.0785]^{12},[3.9685]^{8}\right\} .  \notag \\
d_{ab}^{2\prime } &=&\left\{
[-1.9216]^{8},[-0.2570]^{4},[0]^{5},[0.1406]^{8}\right\} .  \notag
\end{eqnarray}%
In these expressions the superscripts denote the multiplicity of the number
in square brackets.

Thus the algorithm of Ref.~\cite{gudkov02} works for this non-isomorphic
isospectral pair. The dynamical algorithm also distinguishes successfully
some pairs of graphs that have both identical degree distributions and
identical Laplacian spectra\footnote{%
For example, the two graphs of example 2.7 of Russell Merris, Laplacian
Graph Eigenvectors, Linear Algebra and its Applications, v. 278, 1998, pp.
221-236, are distinguished successfully.}.

\begin{figure}[b]
\setlength{\unitlength}{1cm}
\begin{picture}(3,3)

\put(0,0){\circle*{0.1}} \put(0,3){\circle*{0.1}}
\put(1.5,1.5){\circle*{0.1}} \put(3,0){\circle*{0.1}}
\put(3,3){\circle*{0.1}}

\put(1.5,3){\circle*{0.1}} \put(0,1.5){\circle*{0.1}}
\put(3,1.5){\circle*{0.1}} \put(1.5,0){\circle*{0.1}}

\put(0,1.5){\oval(.7,3)[l]} \put(1.5,1.5){\oval(.7,3)[l]}
\put(3,1.5){\oval(.7,3)[l]} \put(1.5,0){\oval(3,.7)[t]}
\put(1.5,1.5){\oval(3,.7)[t]} \put(1.5,3){\oval(3,.7)[t]}

 \put(0,3){\line(1,0){3}}
 \put(0,0){\line(1,0){3}}
  \put(0,0){\line(0,1){3}}
 \put(3,0){\line(0,1){3}}
  \put(1.5,0){\line(0,1){3}}
 \put(0,1.5){\line(1,0){3}}

\end{picture}
\caption{$L_{2}(3)~(9,4,1,2)$}
\label{fig:latticegraph_L2}
\end{figure}
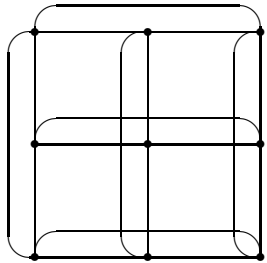

It has been known for decades, however, that certain classes of graphs are
difficult to distinguish by elementary methods. \ An important intransigent
class is the so-called \textquotedblleft strongly regular graphs" (SRG's)%
\cite{west00,brouwer89}. A SRG with parameters $\left( N,k,\lambda ,\mu
\right) $ is a graph with $N$ vertices in which each vertex has $k$
neighbors, each pair of adjacent vertices has $\lambda $ neighbors in
common, and each pair of non-adjacent vertices has $\mu $ neighbors in
common. \ An example known as $L_{2}(3)$ with parameters $(9,4,1,2)$ is
shown in Fig.~\ref{fig:latticegraph_L2}. \ Many pairs of nonisomorphic SRGs
with the same parameter set are known.\footnote{%
A list of all known SRGs with $n<100$ is given at the web site
http://www.cs.uwa.edu.au/$\sim $gordon/remote/srgs/.} \

The adjacency matrix $A$ of a SRG has some interesting algebraic properties.
\ For a general graph, the $(a,b)$ entry of $A^{2}$ is the number of
vertices adjacent to both $a$ and $b.$ \ For SRGs, this number is $\left(
A^{2}\right) _{ab}=k$ if $a=b,$ $\left( A^{2}\right) _{ab}=\lambda $ if $a$
is adjacent to $b,$ and $\left( A^{2}\right) _{ab}=$ $\mu $ if $a$ is not
adjacent to $b.$ \ Hence
\begin{equation*}
A^{2}=kI+\lambda A+\mu (J-I-A),
\end{equation*}%
where $I$ is the identity matrix and $J$ is the matrix consisting entirely
of $1$'s. $J^{2}=NJ.$ \ \ $A$ and $J$ also have the properties that%
\begin{equation*}
AJ=JA=kJ.
\end{equation*}%
This follows from the fact that multiplication of a matrix by $J$ has the
effect of adding the rows or columns of the matrix. \ For $A,$ this sum is
just the number of neighbors. \ \ The Laplacian matrix for a SRG is%
\begin{equation*}
L=kI-A
\end{equation*}%
\ \

These equations show that the three matrices $\left\{ I,J,L\right\} $ form a
commutative, associative algebra. \ The elements of the algebra have the
form $R=$ $fI+gJ+hL,$ where $f,g$ and $h$ are real numbers. \ The
multiplication rule is:
\begin{eqnarray*}
RR^{\prime } &=&R^{\prime }R=\left( fI+gJ+hL\right) \left( f^{\prime
}I+g^{\prime }J+h^{\prime }L\right) \\
&=&\left\{ ff^{\prime }-[k^{2}-k(\lambda -\mu +1)+\mu ]hh^{\prime }\right\}
I+(fg^{\prime }+gf^{\prime }+Ngg^{\prime }+\mu hh^{\prime })J \\
&&+\left[ fh^{\prime }+hf^{\prime }+(2k+\mu -\lambda )hh^{\prime }\right] L
\end{eqnarray*}%
\

\ The structure of the algebra is therefore independent of the precise form
of the $L$ matrix, depending only on the $\left( N,k,\lambda ,\mu \right) $
parameters.

We now consider two non-isomorphic graphs characterized by Laplacian
matrices $L$ and $L^{\prime }$ that share a set $\left( N,k,\lambda ,\mu
\right) .$ \ Let the corresponding dot product matrices be $S=\exp (2LT)$
and $S^{\prime }=\exp (2L^{\prime }T).$ \ Since $L$ and $L^{\prime }$ define
the same algebra, we have%
\begin{equation*}
S\left( N,k,\lambda ,\mu ,T\right) =f\left( N,k,\lambda ,\mu ,T\right)
I+g\left( N,k,\lambda ,\mu ,T\right) J+h\left( N,k,\lambda ,\mu ,T\right) L
\end{equation*}%
and \
\begin{equation*}
S^{\prime }\left( N,k,\lambda ,\mu ,T\right) =f\left( N,k,\lambda ,\mu
,T\right) I+g\left( N,k,\lambda ,\mu ,T\right) J+h\left( N,k,\lambda ,\mu
,T\right) L^{\prime }
\end{equation*}%
where the $f,g,$ and $h$ are definite functions of the parameters --- the
point being that the functions are the same for the two graphs. \ More
explicitly, we have that%
\begin{equation*}
S=\left(
\begin{array}{ccccccc}
f+g+kh & g-hA_{12} & g-hA_{13} & . & . & . & . \\
g-hA_{21} & f+g+kh & g-hA_{23} & . & . & . & . \\
g-hA_{31} & g-hA_{32} & f+g+kh & . & . & . & . \\
. & . & . & . & . & . & . \\
. & . & . & . & . & . & . \\
. & . & . & . & . & .. & . \\
. & . & . & . & . & . & .%
\end{array}%
\right) .
\end{equation*}%
The same relation holds when $S$ is replaced by $S^{\prime }$ and $A$ is
replaced by $A^{\prime }$. \ In any row or column of $A\,$\ or $A^{\prime },$
precisely $k$ entries are equal to $1$ and $N-k$ entries are equal to $0$. \
It now follows that, considered as a set of numbers, $S$ has $N$ entries
equal to $f+g+kh,$ $Nk$ entries equal to $g-h,$ and $N(N-k-1)$ entries equal
to $g$. \ The same holds true for $S^{\prime }.$ \

The $N^{2}$ squared distances satisfy $\left( d^{2}\right)
_{ab}=S_{aa}+S_{bb}-2S_{ab}=2(f+kh)+2hA_{ab}$, and $\left( d^{2}\right)
_{ab}^{\prime }=S_{aa}^{\prime }+S_{bb}^{\prime }-2S_{ab}^{\prime
}=2(f+kh)+2hA_{ab}^{\prime }$ with $a\not=b.$ For $a=b$, $\left(
d^{2}\right) _{ab}=\left( d^{2}\right) _{ab}^{\prime }=0.$ \ There are $Nk$
nonzero entries of $A_{ab}$ and $A_{ab}^{\prime }$ with $a\not=b.$ Hence,
for both graphs there will be $Nk$ distances $2(f+kh)+2h$, $N(N-k-1)$
distances equal to $2(f+kh),$ and $N$ distances equal to $0.$ \ Thus it is
impossible to distinguish this pair of non-isomorphic graphs by this
simplified algorithm.

\begin{figure}[b!]
\setlength{\unitlength}{0.3 cm}
\begin{picture}(15,10)(-2,0)

\multiput(0,0)(2.5,0){5}{\line(0,1){10}}
\multiput(0,0)(0,2.5){5}{\line(1,0){10}} \large{

\put(1,1){4}
 \put(3.5,1){3}
  \put(6,1){2}
 \put(8.5,1){1}
\put(1,3.5){3}
 \put(3.5,3.5){4}
  \put(6,3.5){1}
 \put(8.5,3.5){2}
 \put(1,6){2}
 \put(3.5,6){1}
  \put(6,6){4}
 \put(8.5,6){3}

 \put(1,8.5){1}
 \put(3.5,8.5){2}
  \put(6,8.5){3}
 \put(8.5,8.5){4}

}

\end{picture}
\setlength{\unitlength}{0.3 cm}
\begin{picture}(15,10)(-5,0)

\multiput(0,0)(2.5,0){5}{\line(0,1){10}}
\multiput(0,0)(0,2.5){5}{\line(1,0){10}} \large{

\put(1,1){4}
 \put(3.5,1){1}
  \put(6,1){2}
 \put(8.5,1){3}
\put(1,3.5){3}
 \put(3.5,3.5){4}
  \put(6,3.5){1}
 \put(8.5,3.5){2}
 \put(1,6){2}
 \put(3.5,6){3}
  \put(6,6){4}
 \put(8.5,6){1}

 \put(1,8.5){1}
 \put(3.5,8.5){2}
  \put(6,8.5){3}
 \put(8.5,8.5){4}

}

\end{picture}
\caption{Latin squares $L_3(4)~(16,9,4,6)$\protect\cite{denk74}}
\label{fig:LG_L4}
\end{figure}

To extend this to the actual algorithm of Gudkov \textit{et al}., we examine
their general equations of motion (\ref{eq:motion}). \ From a SRG defined by
$L,$ these equations produce a matrix $X(T)$ which starts life as $X(t=0)=I.$
\ Any numerical solution of the differential equation is simply a finite
sequence of matrix multiplications and additions. \ All such operations
belong to the algebra defined by $I,J,L.$ \ The dot product matrix $S$ also
belongs to this algebra. \ Thus the result is again characterized by only
three numbers $f,g,h.$ \ For a different SRG defined by $L^{\prime }$ but
with the same parameter set, the algorithm produces the same $f,g,h,$ since
the algebra for the two graphs is the same. \ The above argument for the
simplified algorithm then goes through without further modification.

To illustrate the breakdown of the algorithm, we shall apply it to the
smallest pair of non-isomorphic SRGs. \ These are the \textquotedblleft
Latin square\textquotedblright\ graphs with size $N=16.$ Latin squares are
two-dimensional $M\times M$ arrays of the numbers $1$ to $M$, arranged so
that in each row and column no number is repeated. \ Two examples are shown
in Fig.~\ref{fig:LG_L4}. \ Latin square graphs are constructed from Latin
squares as follows: \ Given a Latin square of order $M$, the vertices are
the $N=M^{2}$ cells. \ Two vertices are adjacent if they lie in the same row
or column or if they share the same integer label.

We use the following non-harmonic potential $U_{1}$ to calculate normalized $%
X(T)$ and $d_{ab}^{2}$ again using the first-order Euler algorithm for the
two non-isomorphic Latin square graphs drawn from Fig.~\ref{fig:LG_L4}.
\begin{equation*}
U_{1}=-A\sum (r_{a}-r_{b})^{2}+B\sum (r_{a}-r_{b})^{4}~,
\end{equation*}
where the sum is over pairs of connected vertices. Again taking time
interval $T=1$ with step of length $0.1$, we obtain the distances $d_{ab}^{2}
$ for $A=1,~B=1$
\begin{eqnarray}
d_{ab}^{2} &=&\left\{ [0]^{16},[1.8641]^{96},[2.3129]^{144}\right\} .  \notag
\\
d_{ab}^{2\prime }&=&\left\{ [0]^{16},[1.8641]^{96},[2.3129]^{144}\right\}.
\notag
\end{eqnarray}%
\ $f^{(\prime )},~h^{(\prime )}$ are computed from $d_{ab}^{2(\prime )}$
\begin{eqnarray}
f &=&f^{\prime }=-1.0876  \notag \\
h &=&h^{\prime }=0.2244  \notag
\end{eqnarray}%
$g=$ $g^{\prime },$ but these quantities do not affect the distances. \ We
have also verified that the two sets are identical at each of the discrete
time steps. \ As one would expect, the multiplicity of each distinct
distance depends only on $N$ and $k$ and are independent of time.

\ For completeness we tried a different non-harmonic attractive(repulsive)
potential $U_{1}(\vec{r}_{a}-\vec{r}_{b})(U_{2}(\vec{r}_{a}-\vec{r}_{b}))$
whose force is expressed as%
\begin{eqnarray}
\vec{F}_{1a} &=&-\nabla _{\vec{r}_{a}}U_{1}(\vec{r}_{a}-\vec{r}_{b})={\frac{%
\vec{r}_{a}-\vec{r}_{b}}{1+|\vec{r}_{a}-\vec{r}_{b}|^{3}}}.  \notag \\
\vec{F}_{1a} &=&-\vec{F}_{2a}.  \notag
\end{eqnarray}%
Using this potential and still taking $T=1$, we obtain the distances
\begin{eqnarray}
d_{ab}^{2} &=&\left\{ [0]^{16},[1.4991]^{96},[2.4486]^{144}\right\} .  \notag
\\
d_{ab}^{2\prime } &=&\left\{ [0]^{16},[1.4991]^{96},[2.4486]^{144}\right\} .
\notag
\end{eqnarray}%
with $f^{(\prime )},~h^{(\prime )}$
\begin{eqnarray}
f &=&f^{\prime }=-3.5482  \notag \\
h &=&h^{\prime }=0.4748  \notag
\end{eqnarray}%
Thus, using a non-harmonic potential does not enable the dynamical algorithm
to distinguish these graphs.

Because we have identified two non-isomorphic graphs that result in the same
list of distances, we have disproved the conjecture of Ref.~\cite{gudkov02}.
There are an infinite number of nonisomorphic pairs of SRGs with identical
parameter sets, so the number of counterexample pairs is infinite.

\textbf{Quantum algorithms. \ }The same argument can also be used to show
that a simple one-particle quantum random walk algorithm also fails to solve
the GI problem. \ Consider the vertices of the graph as states $\left\vert
j\right\rangle $ in a Hilbert space. \ The Hamiltonian for the walk is%
\begin{equation*}
H=-\sum_{ab}A_{ab}c_{a}^{\dag }c_{b},
\end{equation*}%
where the operator $c_{a}^{\dag }c_{b}$ is defined by $\left\langle
i\left\vert c_{a}^{\dag }c_{b}\right\vert j\right\rangle =\delta _{ia}\delta
_{bj}.$ \ In physics terms, this is simply a tight-binding model with bonds
on the vertices of the graph. \ We now consider $N$ possible starting
wavefunctions
\begin{equation*}
\left\vert \psi _{1}(t=0)\right\rangle =\left\vert 1\right\rangle
,~\left\vert \psi _{2}(t=0)\right\rangle =\left\vert 2\right\rangle ,~...
\end{equation*}%
and we evolve these forward in time according to the usual time-dependent
Schroedinger equation%
\begin{equation}
i\frac{d\left\vert \psi \right\rangle }{dt}=H\left\vert \psi \right\rangle
\label{eq:schroedinger}
\end{equation}%
for a time $T.$ \ We then compute the $N\times N$ matrix of overlaps%
\begin{equation*}
O_{ij}=\left\langle \psi _{i}(0)|\psi _{j}(T)\right\rangle .
\end{equation*}%
We might then conjecture that this matrix, considered as a set of complex
numbers, is different for non-isomorphic graphs. It can distinguish
isospectral graphs, since \ it uses information about the eigenvectors of $H$%
, not just information about the spectrum. \ However, our algebraic argument
is trivially extended to the now complex algebra defined by $I,J\,,L.$ \ For
non-isomorphic SRG's with identical parameter sets, we again find that the
two matrices of overlaps are the same after a rearrangement, so the
conjecture is invalid.

Though the two strongly regular Latin square graphs of Fig.~\ref{fig:LG_L4}
are not distinguished by the algorithm of Ref.~\cite{gudkov02}, they can be
distinguished in polynomial time by using a construction proposed by Rudolph~%
\cite{rudolph02}, in which the original adjacency matrix is interpreted as
the Hamiltonian of a tight-binding model. The original graph with $N$
vertices describes the possible states (positions) of a single particle. \
However, we can also consider the quantum-mechanical motion of many
particles on the same graph. Rudolph uses symmetric wavefunctions but
forbids double occupancy of any site, corresponding to a hard-core boson
model. Rudolph showed that the spectra of the 3-particle matrices obtained
from two non-isomorphic regular graphs with identical single-particle
spectra are different, demonstrating that the multiparticle construction
does increase the power of the algorithm to distinguish similar graphs.

Here we combine Rudolph's multiparticle construction with the dynamical
algorithm for wavefunction overlaps; this hybrid algorithm has the advantage
that it can distinguish nonisomorphic strongly regular graphs using the
2-particle matrices, as opposed to 3-particle matrices needed if the matrix
eigenvalues are examined. In addition to Rudolph's original case of
hard-core bosons, we also examine non-hard-core bosons and noninteracting
spinless fermions.

For bosons we consider a simple Hubbard Hamiltonian~\cite{hubbard63}, in
which each boson can hop between two vertices if and only if the vertices
are connected by an edge, and in addition there is an energy cost $U$ if two
bosons are on the same site.
Using as a basis the states $\left\vert {ij}\right\rangle $ with particles
at vertices $i$ and $j$, the matrix $K$ for the level $2$ (i.e.
two-particle) graph for bosons is specified by the Hamiltonian matrix
elements
\begin{eqnarray}
K_{ij,~kl}^{B} &\equiv &-\left\langle ij\left\vert H\right\vert
kl\right\rangle  \notag \\
&=&\delta _{il}A_{kj}+\delta _{jk}A_{il}+\delta _{ik}A_{jl}+\delta
_{jl}A_{ik}\qquad \qquad \mathrm{if~}i\neq j{~\mathrm{and~}}k\neq l  \notag
\\
&=&U\delta _{ik}\qquad \qquad \qquad \qquad \qquad \qquad \qquad \qquad
\mathrm{if~}i=j~\mathrm{and~}k=l  \notag \\
&=&{\frac{1}{\sqrt{2}}}(\delta _{il}A_{kj}+\delta _{jk}A_{il}+\delta
_{ik}A_{jl}+\delta _{jl}A_{ik})\qquad \mathrm{if~}i=j{~\mathrm{or~}}k=l,
\notag
\end{eqnarray}%
where for bosonic statistics we have $N(N+1)/2$ initial two-particle states
\begin{equation*}
\left\vert ij\right\rangle =\left\vert 11\right\rangle ,\left\vert
12\right\rangle ,\left\vert 13\right\rangle ,...\left\vert 1N\right\rangle
,\left\vert 22\right\rangle ,\left\vert 23\right\rangle ,\left\vert
24\right\rangle ...,\left\vert NN\right\rangle .
\end{equation*}%
In the hard core limit $U\rightarrow \infty $ the basis states with
doubly-occupied sites can be ignored, the Hilbert space has $N(N-1)/2$
dimensions, and the Hamiltonian matrix elements are
\begin{eqnarray}
K_{ij,~kl}^{HCB} &\equiv &-\left\langle ij\left\vert H\right\vert
kl\right\rangle  \notag \\
&=&\delta _{il}A_{kj}+\delta _{jk}A_{il}+\delta _{ik}A_{jl}+\delta
_{jl}A_{ik}~,  \notag
\end{eqnarray}%
where now we require $i\neq j$ and $k\neq l$.

For Fermi statistics, since two fermions cannot occupy the same vertex, the
Hilbert space has $N(N-1)/2$ dimensions, and we can choose the basis states $%
\left\vert ij\right\rangle ,~i\not=j$. The Hamiltonian matrix elements are
\begin{eqnarray}
K_{ij,~kl}^{F} &=&\delta _{il}A_{kj}+\delta _{jk}A_{il}-\delta
_{ik}A_{jl}-\delta _{jl}A_{ik}.  \notag
\end{eqnarray}

Thus we extend an adjacency matrix of rank $N$ to a matrix $K$ of higher
rank, either $N(N+1)/2$ (for non-hard-core bosons) or $N(N-1)/2$ (for
hard-core bosons and for fermions). Technically, except for hard core
bosons, the matrix $K$ is not an adjacency matrix since it has elements
other than $0$ and $1$.\ For fermionic statistics, some entries of $K$ are
equal to $-1$, while for soft-core bosons some entries are equal to $\sqrt{2}
$ and others to $U$. $K$ can be pictured basically as a matrix in which
every off-diagonal element represents the probability amplitude for
particles to hop from one state to another state.\ Accordingly, the
information of the adjacency matrix $A$ is embedded in the corresponding $K$.

Again we evolve forward the initial two-particle states in time according to
the quantum mechanical evolution~Eq.~(\ref{eq:schroedinger}) and ask whether
any pair of nonisomorphic graphs has two distinct sets of $O_{ij}(T)$.\ The
test of isomorphism is whether the sets $O_{ij}$ and $O_{ij}^{\prime }$ are
the same after rearranging. \ It is of course possible to order these sets
first by their real parts and then by their imaginary parts. \ A simple way
to compare the ordered sets is to compute the numbers $R$ and $I$ defined as
\begin{equation}
R(T)=\sum_{i,j}|\mathrm{{Re}\widetilde{O}_{ij}(T)-{Re}\widetilde{O}%
_{ij}^{\prime }(T)|}  \label{eq:Requation}
\end{equation}%
\begin{equation}
I(T)=\sum_{i,j}|\mathrm{{Im}\widehat{O}_{ij}(T)-{Im}\widehat{O}_{ij}^{\prime
}(T)|}  \label{eq:Iequation}
\end{equation}%
where $\widetilde{O}_{ij},~\widetilde{O}_{ij}^{\prime }$ are the elements of
$O_{ij},~O_{ij}^{\prime }$ ordered by their real parts, while $\widehat{O}%
_{ij},~\widehat{O}_{ij}^{\prime }$ are ordered by their imaginary parts.\ If
either $R(T)$ or $I(T)$ is nonzero, the graphs are non-isomorphic. The
numbers $R$ and $I$ are sufficient to distinguish the non-isomorphic graphs
in this paper.
\begin{figure}[tbp]
\epsfig{figure=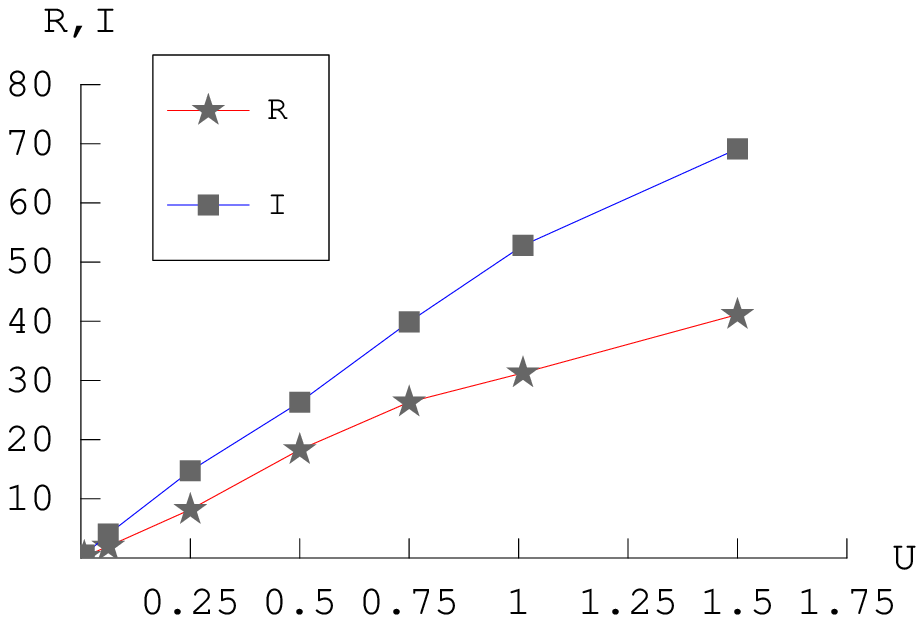,width=4.5 in} \caption{{\protect\small
{Variation of the numbers R and I (defined in
Eqs.~\ref{eq:Requation} and \ref{eq:Iequation}) as a function U
(potential) for the two non-isomorphic Latin square graphs with
N=16. As U goes to zero, R and I vanish.}}} \label{fig:Uplot}
\end{figure}

To probe this approach, we again take pairs of non-isomorphic strongly
regular graphs, two examples being the pairs of non-isomorphic Latin square
graphs in Figs.~\ref{fig:LG_L4} and~\ref{fig:LG_L5}. For each graph in a
given pair, we find numerically the eigenvectors and eigenvalues of $K$ and
use them to calculate the $O_{ij},~O_{ij}^{\prime }$ and then $R(T)$ and $%
I(T)$ for $T=1$. The qualitative behavior does not depend on the choice of $T
$. Table~\ref{tab:results1} shows that for all these pairs of graphs both $R$
and $I$ vanish for noninteracting bosons, but both $R$ and $I$ are nonzero
for hard core bosons and for noninteracting fermions. We find also that $R$
and $I$ are nonzero for graphs with nonzero but finite $U$. \ An
illustration of this is shown in Fig.~\ref{fig:Uplot} the pair with $N=16.$
\ Note that the linear term in $U$ is sufficient to distinguish these
graphs.\

\begin{table}[tbp]
\center
\begin{tabular}{|c|c|c|c|}
\hline
graph specification & fermions & noninteracting bosons & hard-core bosons \\
\hline
(16,9,4,6) & R=1.37 & R=0 & R=110.66 \\
& I=3.01 & I=0 & I=81.53 \\ \hline
(25,12,5,6) & R=1.24 & R=0 & R=129.66 \\
& I=1.93 & I=0 & I=198.53 \\ \hline
(26,10,3,4) & R=1.91 & R=0 & R=14.88 \\
& I=0.65 & I=0 & I=23.47 \\ \hline
(28,12,6,4) & R=1.82 & R=0 & R=87.27 \\
& I=1.25 & I=0 & I=95.11 \\ \hline
(29,14,6,7) & R=3.50 & R=0 & R=28.69 \\
& I=3.73 & I=0 & I=42.51 \\ \hline
\end{tabular}%
\caption{Table of results for hybrid dynamical algorithm for pairs of
nonisomorphic strongly regular graphs for non-interacting fermions,
non-interacting bosons, and hard-core bosons. Each pair of graphs has the
same parameters $(N,k,\protect\lambda,\protect\mu)$, where $N$ is the number
of vertices, each vertex has $k$ neighbors, each pair of adjacent vertices
has $\protect\lambda$ neighbors in common, and each pair of non-adjacent
vertices has $\protect\mu$ neighbors in common. The algorithm distinguishes
the nonisomorphic pairs when using either hard-core bosons or fermions, but
not when using noninteracting bosons.}
\label{tab:results1}
\end{table}

Thus, we find that all the non-isomorphic pairs of strongly regular graphs
that we have examined (with size $N=16,~25,~26,~28,~29$) can be
distinguished by using two-fermion and the interacting boson algorithms. \
It is necessary to examine all matrix elements of $K,$ not just the spectrum
of $K.$ \ The $K$ matrices of the two non-isomorphic graphs still share the
same set of eigenvalues. \ In physical terms, this is due to the additivity
of energies for non-interacting particles. \  
\begin{figure}[b]
\setlength{\unitlength}{0.28 cm}
\begin{picture}(15,10)(0,0)

\multiput(0,0)(2.5,0){6}{\line(0,1){12.5}}
\multiput(0,0)(0,2.5){6}{\line(1,0){12.5}} \large{

\put(1,1){5}
 \put(3.5,1){1}
  \put(6,1){2}
 \put(8.5,1){3}
 \put(11,1){4}
\put(1,3.5){4}
 \put(3.5,3.5){5}
  \put(6,3.5){1}
 \put(8.5,3.5){2}
 \put(11,3.5){3}
 \put(1,6){3}
 \put(3.5,6){4}
  \put(6,6){5}
 \put(8.5,6){1}
 \put(11,6){2}

 \put(1,8.5){2}
 \put(3.5,8.5){3}
  \put(6,8.5){4}
 \put(8.5,8.5){5}
 \put(11,8.5){1}
 \put(1,11){1}
 \put(3.5,11){2}
  \put(6,11){3}
 \put(8.5,11){4}
 \put(11,11){5}

}

\end{picture}
\setlength{\unitlength}{0.28 cm}
\begin{picture}(15,10)(-4,0)

\multiput(0,0)(2.5,0){6}{\line(0,1){12.5}}
\multiput(0,0)(0,2.5){6}{\line(1,0){12.5}} \large{

\put(1,1){5}
 \put(3.5,1){4}
  \put(6,1){2}
 \put(8.5,1){3}
 \put(11,1){1}
\put(1,3.5){4}
 \put(3.5,3.5){3}
  \put(6,3.5){5}
 \put(8.5,3.5){1}
 \put(11,3.5){2}
 \put(1,6){3}
 \put(3.5,6){5}
  \put(6,6){1}
 \put(8.5,6){2}
 \put(11,6){4}

 \put(1,8.5){2}
 \put(3.5,8.5){1}
  \put(6,8.5){4}
 \put(8.5,8.5){5}
 \put(11,8.5){3}
 \put(1,11){1}
 \put(3.5,11){2}
  \put(6,11){3}
 \put(8.5,11){4}
 \put(11,11){5}}
\end{picture}
\caption{Latin squares $L_{3}(5)~(25,12,5,6)$\protect\cite{denk74}}
\label{fig:LG_L5}
\end{figure}

\textbf{Conclusion. \ }Our work suggests two conjectures for the solution of
the GI problem. \ The first, which we doubt but have not been able to
disprove, is that the two-particle quantum algorithm with computation of
overlaps solves GI. \ A more plausible conjecture is that the motion of $%
N_{p}$ quantum-mechanical fermions or interacting bosons on the graph, where
$N_{p}\sim N,$ say $N_{p}=N/2,$ solves GI. \ From the standpoint of
efficiency, we must distinguish whether the algorithm is run on a classical
or a quantum computer. \ The two-particle algorithms already run in
polynomial time on a classical computer. \ The $O_{ij}$ matrix contains $%
O(N^{4})$ entries, each of which is computed in a time of $O(N^{4})$. \ If
this algorithm is sufficient to distinguish all graphs, then GI is in $P$. \
If, on the other hand, we must consider the motion of $N_{p}$ particles,
where $N_{p}\sim N$, then the time is roughly ${\binom{N}{N_{p}}}\sim {%
\binom{N}{N/2}},$ which is exponential in $N$. \ The question of efficiency
on a quantum computer is more interesting. \ The motion of $N_{p}\sim N$
particles can be mapped onto the Heisenberg model of a spin system at a
fixed magnetization, which is nothing more than the dynamics of $N$ qubits
with a constraint. \ Thus the evolution takes place in polynomial time on a
quantum computer. \ On the other hand, we need an exponential number of
overlaps if a naive algorithm is chosen. \ However, since all we need to do
is ask whether the evolution of the two systems is in some sense similar,
there may exist a preparation that entangles the graphs and a measurement
that captures the needed information. \

In conclusion, we have shown that strongly regular graphs provide a useful
testbed for both classical and quantum dynamical algorithms that aim to
solve the graph isomorphism problem. Pairs of graphs in this class cannot be
distinguished by the classical algorithm of Gudkov \textit{et al}.~\cite%
{gudkov02}, but quantum algorithms combining the dynamical evolution of
Gudkov et al.'s algorithm with a construction of Rudolph~\cite{rudolph02}
using interacting bosons as well as noninteracting fermions can distinguish
pairs of SRGs of order $\leq 29,$ while a two-boson noninteracting algorithm
fails.~\cite{big_k_note} Interesting open questions include whether either a
$N_{p}$-fermion or $N_{p}$-hard-core-boson algorithm solves the GI problem,
and if so, what value of $N_{p}$ suffices, and whether an algorithm with $%
N_{p}\sim N$ can be implemented efficiently on a quantum computer.

We thank Eric Bach and Dieter van Melkebeek for extremely useful
conversations. This research was supported by the NSF QuBIC program under
Grant No. NSF-ITR-0130400.


\begin{thebibliography}{99}
\bibitem{factorization_footnote} Perhaps the best-known example of a problem
believed to be neither in P nor NP-complete is factorization, for which no
classical polynomial algorithm is known but which can be solved in
polynomial time on a quantum computer~\cite{shor97}.
\bibitem{adiabatic_footnote} A well-known quantum algorithm motivated by
physical processes is the adiabatic algorithm proposed by Farhi and
collaborators~\cite{farhi01,childs02,roland02}. The algorithms we examine
are clearly different than the adiabatic algorithm, but the philosophy
underlying the scheme is analogous.
\bibitem{big_k_note} We do not know whether the noninteracting $N_{p}$-boson
algorithm distinguishes SRGs when $N_{p}>2$.
\end{thebibliography}
\end{document}